\documentclass[final]{article}
\usepackage[scale=.75]{geometry}
\usepackage{pdflscape}
\usepackage{amsmath, amssymb, mathbbol, mathtools}
\usepackage{svg}
\usepackage{authblk}
\usepackage[colorlinks=true, allcolors=blue]{hyperref}
\usepackage{caption}
\usepackage{subcaption}
\providecommand{\keywords}[1]{\small	\textbf{\textit{Keywords---}} #1}

\newtheorem{rem}{Remark}

\newcommand*{\di}{\mathrm{d}}
\newcommand*{\?}{\mkern-1mu}

\newcommand*{\St}{\mathbb{X}}
\newcommand*{\s}{s}
\renewcommand*{\S}{S}
\newcommand*{\bs}{\boldsymbol{\s}}
\newcommand*{\bS}{\boldsymbol{\S}}
\newcommand*{\bbS}{\mathbb{\S}}
\newcommand*{\q}{q}

\newcommand*{\R}{\mathbb{R}}

\newcommand*{\B}{\mathcal{B}_{\epsilon}}
\newcommand*{\cP}{\mathcal{P}}
\newcommand*{\U}{\mathcal{U}}
\newcommand*{\bX}{\boldsymbol{X}}
\newcommand*{\bx}{\boldsymbol{x}}
\newcommand*{\by}{\boldsymbol{y}}
\newcommand*{\ind}{\mathbb{1}}
\newcommand*{\bY}{\boldsymbol{Y}}
\newcommand*{\sectionname}{Section}
\renewcommand*{\figurename}{Figure}
\newcommand*{\equationname}{Equation}
\newcommand*{\remarkname}{Remark}

\newcommand{\norm}[1]{\left\lVert #1 \right\rVert}

\title{Partial mean-field model for neurotransmission dynamics}

\author[1]{Alberto Montefusco}
\author[2]{Luzie Helfmann}
\author[3]{Toluwani Okunola}
\author[4]{Stefanie Winkelmann}
\author[5]{Christof Schütte}
\affil[1-5]{Mathematics of Complex Systems, Zuse-Institut Berlin, Takustra\ss e 7, 14195 Berlin, Germany}
\affil[3]{Institute of Mathematics, Technische Universit\"at Berlin, Stra\ss e des 17.~Juni 136, 10623 Berlin, Germany}
\affil[5]{Institute of Mathematics, Freie Universit\"at Berlin, Arnimallee 6, 14195 Berlin, Germany}

\date{\today}

\begin{document}

\maketitle
\begin{abstract}
This article addresses reaction networks in which spatial and stochastic effects are of crucial importance. For such systems, particle-based models allow us to describe all microscopic details with high accuracy. However, they suffer from computational inefficiency if particle numbers and density get too large. 
Alternative coarse-grained-resolution models reduce computational effort tremendously, 
e.g., by replacing the particle distribution by a continuous concentration field governed by reaction-diffusion PDEs.
We demonstrate how models on the different resolution levels can be combined into \emph{hybrid models} that seamlessly combine the best of both worlds, describing molecular species with large copy numbers by  macroscopic equations with spatial resolution while keeping the stochastic-spatial particle-based resolution level for the species with low copy numbers. To this end, we introduce a simple particle-based model for the binding dynamics of ions and vesicles at the heart of the neurotransmission process. Within this framework, we derive a novel hybrid model and present results from numerical experiments which demonstrate that the hybrid model allows for an accurate approximation of the full particle-based model in realistic scenarios.
\end{abstract}


\section{Introduction}


Models of spatially well-mixed chemical reaction networks 
have provided a solid foundation for studying molecular and cellular systems; however, the importance of spatial organization in such systems has increasingly been recognized \cite{Winfree2020}. Interacting molecules commonly occur at low copy numbers and move in crowded and diverse environments, so that both stochasticity and spatial resolution play an essential role when modeling biochemical reaction networks. 
Spatial-stochastic simulations have become a prominent tool for understanding how stochasticity at the microscopic level influences the macroscopic behavior of such systems. Recent years have seen increasing interest in particle-based reaction-diffusion models in which all interacting molecules (from ions to entire macromolecules) are single particles diffusing in space, and reactions happen solely if two or more reacting species are in close proximity. Different models and associated simulation environments have been developed: cf.~\cite{Readdy1,Readdy2,tenWolde2017} for examples and \cite{Andrews2018} for an overview. Moreover, there is an extensive literature on using these particle-based models to describe the interplay between spatial organization and stochasticity \cite{WEILANDT2019355,Lowensohn2022,SCHONEBERG20141042}.

While particle-based models guarantee the level of detail necessary to accurately describe the microscopic dynamics, their simulation typically becomes inefficient (or even practically infeasible) for systems with large copy numbers. 
Likewise, so-called agent-based simulations, which become more and more popular for investigating and understanding cellular systems, require cost-effective simulation tools due to the natural complexity of these systems. 
In general, this leads to a conflict of interest between computational efficiency and biochemical accuracy. 

An alternative to developing high-performance computation methods is to study the systems on a theoretical level by finding macroscopic models which approximate the underlying particle-based dynamics. Such a macroscopic approximation not only allows for more efficient simulations, but also gives us a better understanding of the qualitative and quantitative global features of the system. One approach is to study mean-field approximations which approximate the particle-based dynamics in the limit of large numbers of interacting particles. 
Typically, it is shown that the empirical distribution of the particles converges (for an increasing population size) to a concentration field, and the equations governing the particle-based system give rise to a macroscopic equation for this concentration field, e.g., in terms of reaction-diffusion partial differential equations (PDEs) (cf.~\cite{C03-IMS2020reaction,Monte2023,C03-WS19}) or stochastic PDEs (see the extensive literature on fluctuating hydrodynamics \cite{Changho2017}). However, these approaches replace the microscopic, discrete resolution of the particle-based model \emph{completely} by a continuous field. Additionally, recently, methods have been proposed for seamlessly coupling reaction-diffusion PDEs in one spatial compartment (the ``reservoir'') to particle-based simulations in the compartment of interest \cite{C03-KSNR}. These coupled approaches, however, also do \emph{not} solve the conflict of interest if the reaction network under consideration contains molecular species with large copy numbers \emph{as well as} other species with only a few molecules whose specific spatial positions in the cell play an important role in the reaction process. In this case, one would like to construct \emph{hybrid models} that seamlessly combine the best of both worlds, describing the high-abundant species by a macroscopic equation for its concentration field while keeping the stochastic-spatial particle-based resolution level for the low-copy-number species, without spatially separating the two descriptions. 
An important biochemical reaction network containing both low-abundant and high-abundant species is given by the process of neurotransmission which is summarized in the following. 

\paragraph{Background of neurotransmission dynamics.}

\begin{figure}
    \centering
     \begin{subfigure}[b]{0.45\textwidth}\includegraphics[width=\textwidth]{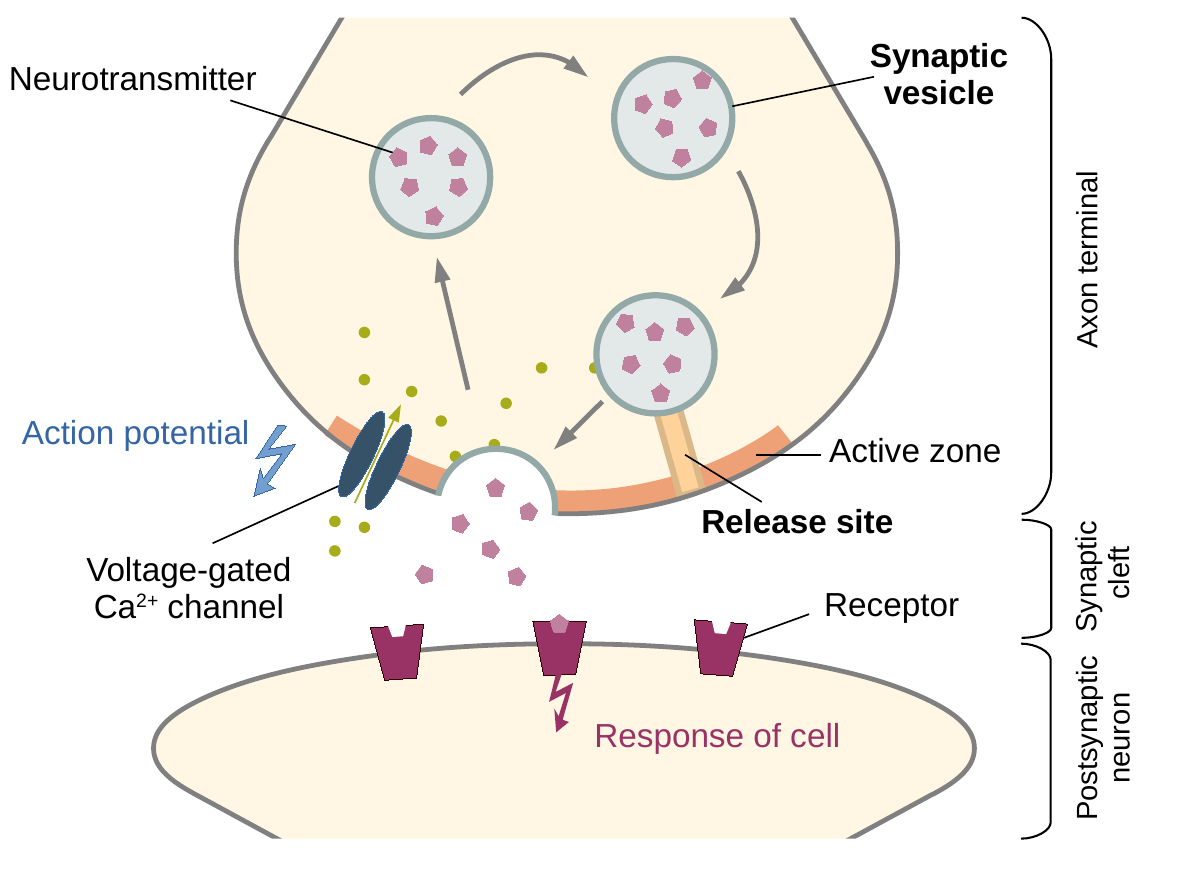} 
    \caption{Illustration of the neurotransmission process. } 
    \label{fig:sketch}
\end{subfigure}
\hfill
\begin{subfigure}[b]{0.45\textwidth}
\includegraphics[width=\textwidth]{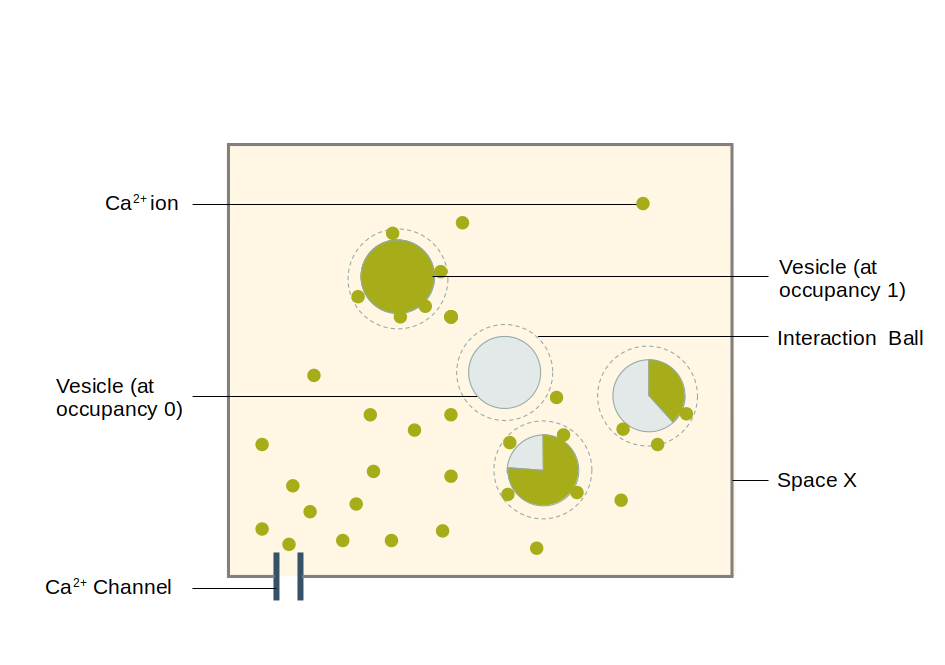}
\caption{Simplified setting in 2d. The vesicles are illustrated as pie charts with olive colour indicating the fraction of occupied ion binding sites and gray colour indicating the unoccupied fraction.}
\label{fig:sketch_simple}
\end{subfigure}
\caption{(a) Illustration of the main aspects of the neurotransmission process between the presynaptic axon terminal of a neuron (upper part with vesicles, ion channel, active zone, and release sites), via the synaptic cleft, to the postsynaptic neuron (lower part with receptors on the surface). (b) Simplified setting used for modelling in this article (square spatial domain $\St$, vesicles with relative occupancy, ions, ion channel). See particle-based model as introduced in \sectionname~\ref{sec:particlebased}.} 
\end{figure}

Neurotransmission is the process of information transfer from one neuron to another (\figurename{}~\ref{fig:sketch}). Within the axon terminal of the presynaptic neuron, the signalling molecules, called \emph{neurotransmitters}, are stored in synaptic vesicles which transport the neurotransmitters to release sites within the so called \textit{active zone} \cite{sudhof2012presynaptic,walter2018vesicle}. Upon stimulation by calcium influx, the vesicles fuse with the membrane to release their content of neurotransmitters into the synaptic cleft where they bind to and activate the receptors of the postsynaptic neuron.
The calcium influx is induced by action potentials which trigger the opening of voltage gated calcium channels \cite{catterall2011voltage}. Calcium ions enter through these channels, diffuse through the axon terminal and bind to the calcium sensors of the vesicles \cite{koh2003synaptotagmin}. The binding of ions to a vesicle increases the probability for the vesicle's fusion to the membrane. It is assumed that there is a maximum number of ions that can attach to a single vesicle (e.g., five ions per vesicle in \cite{kobbersmed2020rapid}). After a fusion event, both the vesicle and the release site undergo a recycling procedure before getting available for reuse \cite{sudhof2004synaptic,ernst2023rate}. 

\paragraph{Modeling neurotransmission dynamics.} Several studies have shown that the process of vesicle fusion and neurotransmitter release is ``stochastic'' in the sense that an arriving action potential does not always elicit fusion \cite{sudhof2004synaptic,allen1994evaluation},  while on the other hand also spontaneous release in the absence of stimuli is possible \cite{goda1997calcium,kavalali2020neuronal}.  This motivates to consider stochastic modeling approaches to describe neurotransmission dynamics. In \cite{kobbersmed2020rapid}, Kobbersmed et al introduce a stochastic vesicle fusion model which describes the dynamics of a set of release sites by a Markovian reaction jump process. The model consists of a set of first-order reactions representing the docking/undocking of a vesicle to the release site, the binding/unbinding of calcium ions, and the fusion event. For some of these reactions, the rates depend on the local calcium concentration which is given as a solution of a PDE taking into account the external calcium concentration and the time point of a stimulus \cite{matveev2002new}. Positions and movement of vesicles and their recycling after fusion, however, are not taken into account; 
instead, it is assumed that there is an infinite supply of vesicles available to all release sites independently of their physical position. This model has been analysed from a mathematical perspective in \cite{ernst2022variance} with a derivation of the characteristic equations for first- and second-order moments of the output current. In \cite{ernst2023rate}, the linear reaction network has been modified by introducing a second-order reaction for the docking of a vesicle to a release site and by adding explicit recovery steps, thereby taking account of the bounded supply of vesicles as well as their recycling.   

\quad

In this article we step beyond the available models and consider a spatially resolved particle-based model for the movement and interactions of vesicles and calcium ions in the axon terminal of the presynaptic neuron. Based on this particle-based model, we construct a hybrid model via a limit process for the high-population species of ions leading to a partial mean-field model coupled to the particle-based model for the low-copy-number species of vesicles. The approximation of the (fully stochastic) dynamics by the hybrid model is well justified by the insight that in the cases of interest there are many more ions present in the axon terminal than release sites or vesicles. The derivation of the model will also show the difficulties and possible pitfalls of hybrid model construction. For the sake of simplicity and transparency, we will restrict both the particle-based and the hybrid model to the core of the neurotransmission process, given by the spatial interaction between the ion field and the stochastic dynamics of the vesicles. Many other processes, like transport through and opening/closing of ion channels, the docking of vesicles to release sites, the vesicle recycling process and the neurotransmitter release itself, are ignored (but can be built it later).

\paragraph{Outline.} At first, we introduce the stochastic particle-based reaction-diffusion model in \sectionname~\ref{sec:particlebased}. The formal derivation of the hybrid model is given in \sectionname~\ref{sec:derivation}. The two models are compared in \sectionname~\ref{sec:NumExp} by means of numerical experiments. Finally, in \sectionname~\ref{sec:Extensions}, we discuss how to expand the models by integrating further aspects of biological detail. 


\section{Particle-based reaction-diffusion model }\label{sec:particlebased}

The particle-based reaction-diffusion model for the spatio-temporal dynamics of vesicles and calcium ions is sketched in \figurename~\ref{fig:sketch_simple}. It will be introduced in the following. 
We will use capital letters, like~$X_i$, for random variables, and small letters, like~$x_i$, for their possible realizations.


\subsection{The configuration space}

The spatial domain is a region~$\St$ within the Euclidean space $(\R^d,\norm{\cdot})$, where the position of each of $m$ vesicles is denoted by~$y_k \in \St$, $k \in \{1, \ldots, m\}$. In the same domain, the position of each of $n$ calcium ions is denoted by $x_i \in \St$, $i \in \{1, \ldots, n\}$, and each ion carries a further internal variable $\s_i \in \{0, \ldots, m\}$ with the following meaning: if $\s_i = k$, the $i$-th ion is bound to the $k$-th vesicle, while $\s_i=0$ means that the $i$-th ion is free/unbound. Each vesicle can bind at most $n_v \coloneqq \lfloor a \, n \rfloor$ ions with a ratio $a \in [0, 1]$.
The configuration space is thus characterized by the triple of vectors $(\bx, \bs, \by) \in \St^n {\times} \bbS_{n, m} {\times} \St^m$, where the space 
\begin{equation}
    \bbS_{n, m} \coloneqq \left\{\bs \in \{0,...,m\}^n \bigm| \sum_{i=1}^n \delta_{\s_i, k} \leq n_v \ \forall k \in \{1, ..., m\}\right\}
\end{equation}
ensures that each vesicle binds no more than $n_v$~ions.

\quad

The particle-based dynamics is given by the stochastic process 
\begin{equation*}
(\bX(t),\bS(t),\bY(t))_{t\geq 0}\in  \St^n {\times} \bbS_{n, m} {\times} \St^m,
\end{equation*}
 where $\bX(t)=(X_i(t))_{i=1,...,n}$ refer to the ion's positions, $\bS(t)=(\S_i(t))_{i=1,...,n}$ give their binding state, and $\bY(t)=(Y_k(t))_{k=1,...,m}$ are the vesicle's positions. 
The dynamics is a superposition of two types of stochastic processes: a diffusive component for the positions of both the vesicles and the unbound ions, and a reaction component for the binding of the ions to the vesicles as well as their unbinding.

\subsection{Position dynamics}

The $k$-th vesicle moves  according to the overdamped Langevin equation
\begin{equation}\label{position_vesicle}
    dY_k(t) = -\biggl(\nabla V\bigl(Y_k(t)\bigr) + \sum_{\ell \neq k} \nabla U\bigl(Y_k(t) - Y_\ell(t)\bigr) \biggr) \, dt  + \sigma^Y dW_k^Y\?(t) ,
\end{equation}
where $\sigma^Y > 0$ is the noise intensity, $(W_k^Y\?(t))_{t\geq 0}$, $k \in \{1, ..., m\}$, are $d$-dimensional independent Wiener processes, $V \colon \St \to \mathbb{R}$ is a potential field, and $U \colon \St \to \mathbb{R}$ generates a short-range repulsion -- for instance given by an exclusion force.

In an analogous but simpler fashion, the position of each unbound ion $i$ (with $\S_i(t)=0$) evolves according to a stochastic process~$X_i$ given by the Brownian motion
\begin{equation}\label{position_ion}
    dX_i(t) = \delta_{\S_i(t),0} \, \sigma^X \, dW_i^X\?(t) \qquad i \in \{1, \ldots n\} ,
\end{equation}
with noise intensity $\sigma^X > 0$ and independent Wiener processes $W_i^X\?$.
A bound ion $i$ is assumed to move with the vesicle $k$ it is attached to (i.e., $X_i(t)=Y_k(t)$ for all times where $\S_i(t)=k$) and only starts moving independently again when it unbinds from it. The spatial trajectories of the ions are thus piecewise continuous with discontinuities restricted to the time points where binding or unbinding occurs.

Both vesicles and ions are restricted to stay in the domain $\St$, which is implemented by reflecting boundary conditions. 

\subsection{Binding and unbinding}\label{sec:binding}

When the $i$-th ion is unbound ($\S_i(t)=0$) and $\epsilon$-close to the $k$-th vesicle, i.e., $X_i(t)\in \B(Y_k(t))$ for 
\begin{equation*}
    \B(y) \coloneqq \bigl\{x \in \St \bigm\vert \norm{x-y}\leq \epsilon\bigr\} ,
\end{equation*}
 then it has a certain probability to bind to that vesicle. 
 We assume that the binding rate only depends on the \emph{relative occupancy} of the particular vesicle, which we define as
\begin{equation}\label{w_k}
     w_k \coloneqq \frac{1}{n_v} \sum_{i=1}^n \delta_{\s_i, k} \in [0, 1] 
\end{equation}
given the binding state $\bS(t) =\bs \in \bbS_{n, m}$. 
The binding rate is thus of the form~$r^+\?(w_k)$ with some function $r^+ \colon [0, 1] \to \R_+$ that we specify later. From the moment~$t$ where the ion becomes bound, it assumes the position~$Y_k(t)$ of the vesicle, such that $X_i(t') = Y_k(t')$ for $t'\geq t$, until it unbinds again. 
Analogously, an ion that is bound to the $k$-th vesicle can unbind from it at rate~$r^-(w_k)$ with $r^- \colon [0, 1] \to \R_+$.  After unbinding, it starts from a new position extracted randomly, according to some distribution~$\mu_k$, inside the ball around $Y_k(t)$. A concrete choice for the distribution is the uniform $\U\bigl(\B\bigl(Y_k(t)\bigr)\bigr)$. We will formulate the generator of this dynamics in \sectionname ~\ref{sec:generator}. 

\paragraph{The binding and unbinding rate functions.} 
It remains to specify the functions~$r^\pm$. Different choices exist in the literature \cite{shomar2017cooperative,reva2021first,matveev2022close} and are based on both theoretical and empirical grounds. The simplest form of the binding rate is given by
\begin{equation}\label{eq:r+uncoop}
    r^+\?(w) = \gamma^+ \, (1-w) , \quad w \in [0,1] 
\end{equation}
with $\gamma^+ > 0$, which decreases linearly with the number of available binding sites. The unbinding rate may in its simplest form be assumed to be a constant
\begin{equation}\label{eq:r-uncoop}
    r^-\?(w) = \gamma^-,
\end{equation}
with $ \gamma^- > 0$,
implying that unbinding is independent from the number of currently bound ions.

A typical feature that emerges from the literature and is supported by experimental evidence is the so-called \emph{cooperativity}: the more ions are bound to the vesicle, the easier for a new ion to bind and the harder for a bound ion to unbind. 
This form of attractive force between calcium ions is modeled in different ways in the literature. 
\begin{itemize}
    \item \cite{shomar2017cooperative} suggests a cooperative binding rate of the form 
    \begin{equation}\label{eq:r+cooperative}
        r^+\?(w) = \gamma^+ (w + \alpha^+)\, (1-w) .
    \end{equation}
    Hence, the binding rate not only decreases with a decreasing number of binding sites $\propto (1-w)$, but also increases with an increasing number of bound ions $\propto (w+\alpha^+)$ because of an attracting interaction between the calcium ions. The additive constant $\alpha^+>0$ ensures that binding is also possible when $w=0$. 
    \item The same work~\cite{shomar2017cooperative} suggests a cooperative unbinding rate of the form 
    \begin{equation}\label{eq:r-cooplinear}
        r^-\?(w) = \gamma^-(1-w+\alpha^-) ,
    \end{equation}
    which makes it linearly harder for the ion to unbind the more ions are already bound. The factor $\alpha^->0$ ensures that unbinding is also possible when $w=1$.
    \item In~\cite{schneggenburger2000intracellular,matveev2022close,reva2021first}, the cooperative unbinding function is assumed to decay  exponentially in $w$ and is of the general form 
    \begin{equation}\label{eq:r-coopexp}
        r^-\?(w) = \gamma^- \beta^{w}
    \end{equation}
     with $\beta\in (0,1)$. Thus, unbinding becomes exponentially harder the more ions are already bound.
\end{itemize}
Numerical experiments to study the dynamics for the different types of rate functions will be given in \sectionname~\ref{sec:NumExp}.

\section{Partial mean-field model}\label{sec:derivation}

When following the detailed trajectories of all particles is either unfeasible or uninteresting, a description of the system in terms of a collective variable may give us the possibility of faster simulations and a better understanding of the qualitative and quantitative global features of the system. Furthermore, when the number of particles is sufficiently large, there is a chance to have a simpler description of the system by reducing the noise -- or part of it -- to some deterministic dynamics.

In our model, we are interested in keeping track of the spatial concentration of unbound calcium ions and the positions and occupancies of all vesicles. The goal is to derive, in the limit where the number of calcium ion is sufficiently large, a PDE for the spatial calcium concentration coupled to an ordinary differential equation (ODE) for the relative occupancy state $w(t)$ of the vesicles, 
while keeping the particle-based resolution for their movement.


For the sake of clarity, we confine the formal derivation of \sectionname{}s~\ref{sec:generator}-\ref{sec:limit} to a simplified setting and focus on what happens around a single vesicle ($m = 1$) with a fixed position $Y_1(t) = y\in \mathbb{X}$ for all~$t$. The configuration space is then $\St^n \times \bbS_n$, where $\bbS_n \coloneqq \bbS_{n, 1}$, with states of the form $(\bx,\bs)$ for the positions and binding states of the $n$ calcium ions. Moreover, we will neglect boundary conditions. The extension to the complete model will be considered in \sectionname ~\ref{sec:hybrid}.

\subsection{Derivation of the generator for the empirical measure}\label{sec:generator}

The central object in our derivation is the empirical measure
\begin{equation}\label{empirical_measure}
    \rho^n_{\bX, \bS}(\di x', \s') \coloneqq \frac1n \sum\limits_{i=1}^n \delta_{X_i}(\di x') \, \delta_{\S_i, \s'} ,
\end{equation}
which counts the relative number of ions in the volume~$\di x'$ and with binding state~$\s'$ (unbound if $\s' = 0$, bound if $\s' = 1$). 
Instead of manipulating the stochastic processes directly, we take a ``weak'' viewpoint and work with their associated infinitesimal generators. In the present section, we derive, from the infinitesimal generator~\eqref{generator} for $(\bX(t), \bS(t))$, the generator~\eqref{measure_generator} for the measure-valued process $\rho^n_{\bX(t), \bS(t)}$. Then, in \sectionname ~\ref{sec:limit}, we look for its deterministic limit as $n \to \infty$, and finally project the dynamics further. The derivations will not be rigorous, but the language will be close to the mathematical formalism that would be necessary for a full proof.

The starting point of the derivation requires the infinitesimal generator for the process $(\bX(t), \bS(t))_{t \geq 0}$. This contains a diffusion component for the positions of the calcium ions, as well as a binding and an unbinding component. Since, in \sectionname ~\ref{sec:limit}, we will perform the limit when the number of ions~$n$ is large, we stress the dependence of the generator on the parameter~$n$ by denoting it as $L^n$. For any observable $f \in C^{2,0}(\St^n {\times} \bbS_n)$, we have
\begin{align}\label{generator}
    \MoveEqLeft (L^n f)(\bx, \bs) \nonumber \\ 
    ={}& \frac{\sigma^2}{2} \sum\limits_{i=1}^n \delta_{\s_i, 0}  \, \Delta_{x_i} f(\bx, \bs) \nonumber \\
    &+ r^+\?(w) \sum\limits_{i=1}^n \ind_{\B(y)}\?(x_i) \, \delta_{\s_i, 0} \, \bigl[f(x_1, ..., x_{i-1}, y, x_{i+1}, ..., x_n, \s_1, ..., 1-\s_i, ..., \s_n) - f(\bx, \bs) \bigr] \nonumber \\
    &+ r^-(w) \sum\limits_{i=1}^n \delta_{\s_i, 1} \, \int_{\St} \mu(\di x'_i) \, \bigl[f(x_1, ..., x_{i-1}, x'_i, x_{i+1}, ..., x_n, \s_1, ..., 1-\s_i, ..., \s_n) - f(\bx, \bs) \bigr] ,
\end{align}
where $w \coloneqq w_1 \in [0,1]$ is a placeholder for the relative occupancy of the single vesicle with fixed position $y$, given by
\begin{equation*}
    w=\frac{1}{\lfloor an \rfloor} \sum_{i=1}^n \delta_{s_i, 1} .
\end{equation*}
The first term in~\eqref{generator} contains the second derivative of the observable~$f$ and corresponds to diffusion of the unbound ions, where $\sigma \coloneqq \sigma^X$ is the noise intensity. The second term refers to the binding of the $i$-th ion (if unbound and $\epsilon$-close to the vesicle), which is placed at the position~$y$ of the vesicle. The third term refers to the unbinding of the $i$-th ion (if bound) and its replacement around the position of the vesicle.
A uniform replacement corresponds to the distribution
\begin{equation*}
    \mu(\di x') = \frac{1}{\lvert\B\rvert} \ind_{\B(y)}(x') \, \di x' .
\end{equation*}

The infinitesimal generator~\eqref{generator} acts on observables for the pair~$(\bx, \bs)$. To operate the passage from $(\bx, \bs)$ to the empirical measure, we apply the generator to observables of the form  
\begin{equation*}
    f(\bx, \bs) = g(\rho^n_{\bx, \bs}) ,
\end{equation*}
for $g\in C^2(\cP(\St {\times} \{0, 1\}))$
which depend on $(\bx, \bs)$ only through the empirical measure. Ideally, we would hope that the function~$L^n f$ also depends on $(\bx, \bs)$ solely through the empirical measure and, as a consequence, we would be able to identify a generator for the Markov process~$\rho^n_{\bS, \bX}$. This will be our case, as we will see in the following.\footnote{In general, one cannot accomplish this procedure so easily, but often can still recover an autonomous equation for the empirical measure in the deterministic limit, namely when $n \to \infty$ \cite[Chapter~5]{KL99}.} The final step is to take the limit of the generator for $\rho^n_{\bX, \bS}$ as $n \to \infty$: in our situation, we obtain another generator which contains only a drift term and thus corresponds to a deterministic PDE.

We thus need to find good observables~$g$ for the empirical measure, such that they fully characterize the generator: the set of observables can be smaller than the domain of the generator, but still has to be big enough.\footnote{According to semigroup theory, a subset of the domain that fully characterizes the generator is a \emph{core} \cite[Chapter~II]{EN06}.} Since the empirical measure is an infinite-dimensional object, it is convenient to consider a finite-dimensional projection by testing it with a finite set of continuous and bounded functions $\phi_\ell \in C_b(\St {\times} \{0, 1\})$, $\ell = 1, ..., \q$:
\begin{equation*}
    \rho \longmapsto \bigl( \langle \phi_1, \rho \rangle, \langle \phi_2, \rho \rangle, \ldots, \langle \phi_\q, \rho \rangle \bigr) 
\end{equation*}
for $\rho \in \cP(\St{\times}\{0, 1\})$, 
where
\begin{equation}\label{def:phi_rho}
    \langle \phi, \rho \rangle \coloneqq \sum\limits_{\s\in\{0, 1\}} \int_{\St} \phi(x, \s) \, \rho(\di x, \s) .
\end{equation}
 The projection onto a finite-dimensional space makes the successive calculations manageable -- these are reduced to ordinary calculus -- without loss of generality, since we consider \emph{all} possible projections, i.e., all possible test functions.

The corresponding simplified observables for the empirical measure are the \emph{cylindrical functions} \cite[Definition~5.1.11]{AGS08}
\begin{equation*}
    g(\rho) \coloneqq \psi\bigl( \langle \phi_1, \rho \rangle, \langle \phi_2, \rho \rangle, \ldots, \langle \phi_\q, \rho \rangle \bigr) ,
\end{equation*}
with $\psi \in C^2(\R^\q)$. We have thus traded a function~$g$ on a infinite-dimensional space for a function~$\psi$ on an Euclidean space. We now make use of the cylindrical functions $g$ and consider the following observables for $(\bx, \bs)$:
\begin{align}\label{observables}
    f(\bx, \bs) = g(\rho^n_{\bx, \bs}) &= \psi\bigl( \langle \phi_1, \rho^n_{\bx, \bs} \rangle, \langle \phi_2, \rho^n_{\bx, \bs} \rangle, \ldots, \langle \phi_\q, \rho^n_{\bx, \bs} \rangle \bigr) \\
    &= \psi\biggl( \frac1n \sum\limits_{i=1}^n \phi_1(x_i, \s_i), \frac1n \sum\limits_{i=1}^n \phi_2(x_i, \s_i), \ldots, \frac1n \sum\limits_{i=1}^n \phi_\q(x_i, \s_i) \biggr) . \nonumber
\end{align}
Since these functions depend on $(\bx, \bs)$ only through the empirical measure, they do not depend on the permutations of particles, namely are invariant under any permutation that is performed in both $\bx$ and $\bs$. The goal then is to show that the function $L^n f$ also depends on $(\bx, \bs)$ only through the empirical measure.

As a result of the application of the generator~\eqref{generator} to the observables~\eqref{observables}, we obtain
\begin{align}\label{generator_final}
    \MoveEqLeft (L^n f)(\bx, \bs) \\
    ={}& \frac{\sigma^2}{2} \int_{\mathbb{X}} \rho^n_{\bx, \bs}(\di x', 0) \sum\limits_{\ell=1}^\q \partial_\ell\psi\bigl(\langle \phi_1, \rho^n_{\bx, \bs} \rangle, ...\bigr) \, \Delta_1\phi_\ell(x', 0) \nonumber \\
    &+ \frac{\sigma^2}{2n} \int_{\mathbb{X}} \rho^n_{\bx, \bs}(\di x', 0) \sum\limits_{k=1}^\q \sum\limits_{\ell=1}^\q \partial^2_{k \ell} \psi\bigl(\langle \phi_1, \rho^n_{\bx, \bs} \rangle, ...\bigr) \, \nabla_1 \phi_k(x', 0) \cdot \nabla_1 \phi_\ell(x', 0) \nonumber \\
    &+ n \int_{\mathbb{X}} \rho^n_{\bx, \bs}(\di x', 0) \, \ind_{\B(y)}(x') \, r^+\?(w) \biggl[ \psi\Bigl(\langle \phi_1, \rho^n_{\bx, \bs} \rangle - \frac1n \phi_1(x', 0) + \frac1n \phi_1(y, 1), ...\Bigr) - \psi\bigl(\langle \phi_1, \rho^n_{\bx, \bs} \rangle, ...\bigr) \biggr] \nonumber \\
    &+ n \iint_{\mathbb{X}^2} \rho^n_{\bx, \bs}(\di x', 1) \, \mu(\di x'') \, r^-\?(w) \biggl[ \psi\Bigl(\langle \phi_1, \rho^n_{\bx, \bs} \rangle - \frac1n \phi_1(x', 1) + \frac1n \phi_1(x'', 0), ...\Bigr) - \psi\bigl(\langle \phi_1, \rho^n_{\bx, \bs} \rangle, ...\bigr) \biggr] , \nonumber
\end{align}
where $\nabla_1$ and $\Delta_1$ act on the first variable, and we clearly need that $\phi_\ell \in C^{2,0}(\St {\times} \{0, 1\})$ for every~$\ell$. The placeholder~$w$ is now interpreted in terms of~$\rho^n_{\bx, \bs}$ as
\begin{equation*}
    w=\frac{n}{\lfloor an \rfloor}\int_{\St} \rho^n_{\bx,\bs}(\di x',1)=\frac{n}{\lfloor an \rfloor} \rho^n_{\bx,\bs}(\{y\},1)
\end{equation*}
Note, indeed, that all bound ions are placed at~$y$, and therefore the measure~$\rho(\cdot, 1)$ concentrates fully on~$y$. 
The complete steps of the calculations are shown in Appendix~\ref{appendix}.
As hoped for, the generator depends on $(\bx, \bs)$ only through the empirical measure.

To write down the final expression of the generator for the process~$\rho^n_{\bX, \bS}$ of empirical measures, we need the derivatives of the cylindrical functions~$g(\rho)$, which we compute via the chain rule. Since the functional derivative of the linear function $\rho \mapsto \langle \phi, \rho \rangle$ (see~\eqref{def:phi_rho}) is simply $\phi \in C_b(\St{\times}\{0,1\})$, we have
\begin{subequations}
\begin{align}
    g'(\rho) &= \sum\limits_{\ell=1}^q \partial_\ell\psi\bigl(\langle \phi_1, \rho \rangle, ..., \langle \phi_\q, \rho \rangle \bigr) \, \phi_\ell &&\in C_b(\St {\times} \{0, 1\}) , \\
    g''(\rho) &= \sum\limits_{k=1}^\q \sum\limits_{\ell=1}^\q \partial^2_{k \ell} \psi\bigl(\langle \phi_1, \rho \rangle, ..., \langle \phi_q, \rho \rangle \bigr) \, \phi_k \otimes \phi_\ell &&\in C_b(\St {\times} \{0, 1\}) \otimes C_b(\St {\times} \{0, 1\}) .
\end{align}
\end{subequations}
Given these derivatives, we find an expression that can be written fully in terms of the function~$g$,
\begin{align*}
    (L^n f)(\bx, \bs) ={}& \frac{\sigma^2}{2}  \int_{\mathbb{X}} \rho^n_{\bx, \bs} (\di x', 0) \, \Delta_1(g'(\rho^n_{\bx, \bs}))(x', 0) + \frac{\sigma^2}{2n} \int_{\mathbb{X}} \rho^n_{\bx, \bs}(\di x', 0)  \, (\nabla_1 \cdot \nabla_3) (g''(\rho^n_{\bx, \bs}))(x', 0, x', 0) \nonumber \\
    &+ n \int_{\B(y)} \rho^n_{\bx, \bs}(\di x', 0) \, r^+\?(w) \, \biggl[ g\Bigl(\rho^n_{\bx, \bs} - \frac1n \delta_{x'} \, \delta_0 + \frac1n \delta_y \, \delta_1\Bigr) - g(\rho^n_{\bx, \bs}) \biggr] \nonumber \\
    &+ n \iint_{\mathbb{X}^2} \rho^n_{\bx, \bs}(\di x', 1) \, \mu(\di x'') \ r^-\?(w) \biggl[ g\Bigl(\rho^n_{\bx, \bs} - \frac1n \delta_{x'} \, \delta_1 + \frac1n \delta_{x''} \, \delta_0\Bigr) - g(\rho^n_{\bx, \bs}) \biggr] ,
\end{align*}
and thus, after replacing $\rho^n_{\bx, \bs}$ by $\rho$, arrive at the generator $Q^n$ for the Markov process~$\rho^n_{\bX, \bS}$:
\begin{align}
    (Q^n g)(\rho) ={}& \frac{\sigma^2}{2}  \int_{\mathbb{X}} \rho(\di x, 0) \, \Delta_1(g'(\rho))(x, 0) + \frac{\sigma^2}{2n} \int_{\mathbb{X}} \rho(\di x, 0)  \, (\nabla_1 \cdot \nabla_3) (g''(\rho))(x, 0, x, 0) \nonumber \\
    &+ n \int_{\B(y)} \rho(\di x, 0) \, r^+\?(w) \, \biggl[ g\Bigl(\rho - \frac1n \delta_x \, \delta_0 + \frac1n \delta_{y} \, \delta_1\Bigr) - g(\rho) \biggr] \nonumber \\
    &+ n \iint_{\mathbb{X}^2} \rho(\di x, 1) \, \mu(\di x') \ r^-\?(w) \biggl[ g\Bigl(\rho - \frac1n \delta_x \, \delta_1 + \frac1n \delta_{x'} \, \delta_0\Bigr) - g(\rho) \biggr] .
 \label{measure_generator}
\end{align}
This is the generator of an infinite-dimensional measure-valued Markov process and, for us, represents the starting point to derive the partial mean-field model. The expression is very general and accommodates measures~$\rho$ that do not have any Lebesgue density -- like for instance Dirac measures.

Before performing the last step and sending $n \to \infty$ in~\eqref{measure_generator}, we examine the various terms in the generator and highlight their contribution to the measure-valued process in the following remark.

\begin{rem}
The operator~$Q^n$ contains three types of terms \cite{daD93}:
\begin{itemize}
    \item A first-derivative term, which corresponds to a drift. This is the contribution
    \begin{equation*}
        \frac{\sigma^2}{2} \int_{\St} \rho(\di x, 0) \, \Delta_1 \varphi(x, 0)
    \end{equation*}
    for a test function~$\varphi \in C^{2,0}(\St {\times} \{0, 1\})$. This term alone is the weak form of a parabolic diffusion equation for~$\rho(\cdot, 0)$, which in strong form would be
    \begin{equation*}
        \frac{\sigma^2}{2} \Delta_1 \rho(x, 0) .
    \end{equation*}
    \item A second-derivative term which corresponds to a stochastic diffusion and has one order in~$n$ less than the drift one. The underlying bilinear form (a diffusion tensor) is the integral form
    \begin{equation*}
        D(\varphi_1, \varphi_2) \coloneqq \frac{\sigma^2}{2n} \int_{\St} \rho(\di x, 0) \, \nabla_1 \varphi_1(x, 0) \cdot \nabla_1 \varphi_2(x, 0) .
    \end{equation*}
    To display a strong form, we can perform a formal integration by parts and obtain
    \begin{equation*}
        D(\varphi_1, \varphi_2) = - \frac{\sigma^2}{2n} \int_{\St} \nabla_1 \cdot \bigl(\rho(x, 0) \, \nabla_1 \varphi_1(x, 0) \bigr) \, \varphi_2(x, 0) \, \di x .
    \end{equation*}
    The ``square root'' of the diffusion matrix is the noise intensity that would appear in the corresponding stochastic partial differential equation, where it acts on the space-time white noise (cf.~\cite{dsD96}).
    \item A finite-difference term, which corresponds to jumps of the form
    \begin{align*}
        \rho \ \longmapsto \ \rho - \frac1n \delta_x \, \delta_0 + \frac1n \delta_y \, \delta_1 &\qquad \text{with transition rate density} \quad n \, r^+\?(w) \, \rho(\di x, 0) , \text{ and} \\
        \rho \ \longmapsto \ \rho - \frac1n \delta_x \, \delta_1 + \frac1n \delta_{x'} \, \delta_0 &\qquad \text{with transition rate density} \quad n \, r^-\?(w) \, \rho(\di x, 1) \, \mu(\di x') . 
    \end{align*}
\end{itemize}
The three terms essentially reflect the features of the original particle-based process~$(\bX, \bS)$: the diffusion of the ions has been translated into a drift of the empirical measure and a lower-order diffusion term; the jumps have remained the same, with rates that are proportional to the empirical measure.
\end{rem}

\subsection{Deterministic limit}\label{sec:limit}

As $n \to \infty$, we expect the process $\rho^n_{\bX, \bS}$ to become more and more deterministic, namely concentrated on a continuous measure-valued trajectory. The trajectory is the solution of a measure-valued PDE. Here we give a heuristic derivation of such a PDE by performing a formal Taylor expansion around~$\frac1n = 0$ of the jump terms in the generator~\eqref{measure_generator}:
\begin{align*}
    n \, \Bigl[ g\Bigl(\rho - \frac1n \delta_x \, \delta_{\s} + \frac1n \delta_{x'} \, \delta_{\s'}\Bigr) - g(\rho) \Bigr] &= n \Bigl[ \psi\Bigl(\langle \phi_1, \rho \rangle - \frac1n \phi_1(x, \s) + \frac1n \phi_1(x', \s'), ...\Bigr) - \psi\bigl(\langle \phi_1, \rho\rangle, ...\bigr) \Bigr] \\
    &= \sum\limits_{\ell=1}^q \partial_\ell \psi\bigl(\langle \phi_1, \rho \rangle, ...\bigr) \, \bigl(\phi_\ell(x', \s') - \phi_\ell(x, \s)\bigr) + o(1)\\
    &= g'(\rho)(x', \s') - g'(\rho)(x, \s) + o(1) .
\end{align*}

Then, we replace the corresponding terms in the generator~\eqref{measure_generator} and obtain, upon sending $n \to \infty$,
\begin{align}
    (Q^n g)(\rho) \to (Q^\infty g)(\rho) ={}& \frac{\sigma^2}{2}  \int_{\mathbb{X}} \rho(\di x, 0) \, \Delta_1(g'(\rho))(x, 0) \nonumber \\
    &+ \int_{\B(y)} \rho(\di x, 0) \, r^+\?(w) \, \bigl( g'(\rho)(y, 1) - g'(\rho)(x, 0) \big) \nonumber \\
    &+ \iint_{\mathbb{X}^2} \rho(\di x, 1) \, \mu(\di x') \ r^-\?(w) \bigl( g'(\rho)(x', 0) - g'(\rho)(x, 1) \bigr) 
\end{align}
with $w= \frac{1}{a}\rho(\{y\},1)$. 
The limit generator~$Q^\infty$ contains only first derivatives of the observables and therefore is the generator of a deterministic (measure-valued) process.\footnote{A generator $L$ of the form $(Lf)(x) = A(x) \cdot f'(x)$ containing only first derivatives of the argument is associated with the deterministic differential equation $\dot x = A(x)$; such a generator is the transpose operator of the operator that generates the Liouville equation.} 
Its paths are the solutions~$(\rho(\cdot; t))_{t\geq 0}$ of the equation
\begin{align}\label{limit}
    \frac{\di}{\di t} \sum\limits_{\s \in \{0, 1\}} \int_{\St} \rho(\di x, \s; t) \, \varphi(x, \s) ={}& \frac{\sigma^2}{2}  \int_{\mathbb{X}} \rho(\di x, 0; t) \, \Delta_1\varphi(x, 0) + \int_{\B(y)} \rho(\di x, 0; t) \, r^+\?(w(t))\, \bigl( \varphi(y, 1) - \varphi(x, 0) \big) \nonumber \\
    &+ \iint_{\mathbb{X}^2} \rho(\di x, 1; t) \, \mu(\di x') \, r^-\?(w(t)) \bigl( \varphi(x', 0) - \varphi(x, 1) \bigr) ,
\end{align}

As a final step, we aim to find the evolution equations for the relative occupancy~$w(t)$
and the concentration of unbound ions. Since we write them in strong form, we define the concentration as the Lebesgue density of~$\rho(\cdot, 0)$:
\begin{equation}
    c(x) \coloneqq \rho(x, 0) .
\end{equation}
If $\mu$ has a Lebesgue density too, we can perform an integration by parts in \eqref{limit} and find
\begin{subequations}\label{cw}
\begin{equation}
    \dot{c}(x; t) = \frac{\sigma^2}{2} \Delta c(x; t) - \ind_{\B(y)}(x) \, r^+\?(w(t)) \, c(x; t) + r^-\?(w(t)) \, \mu(x) \, a \, w(t) . \label{c(t)}
\end{equation}
The equation for~$w(t)$ is recovered from \eqref{limit} by using the concentration property~$w(t) = \frac1a \rho(\{y\}, 1; t)$:
\begin{equation}
    \dot{w}(t) = r^+\?(w(t)) \frac1a \int_{\B(y)} c(x; t) \, \di x - r^-\?(w(t)) \, w(t) . \label{w(t)}
\end{equation}
\end{subequations}

\begin{rem}\label{rem:occu}
\equationname~\eqref{w(t)} marks a crucial step in this derivation. We started with a purely discrete object (the relative occupancy $w_k$ defined in~\eqref{w_k} with a finite state space) and replaced it with the object $w(t)$, which continuously evolves in space $[0,1]$.
The continuous occupancy $w(t)$ is the $n\to \infty$ limit of the sequence of discrete occupancies. This step is based on a \emph{scaling assumption}: When the number $n$ of ions grows, the number of ions that can be bound to an individual vesicle grows as well (its maximum $n_v\coloneqq \lfloor a \, n \rfloor$ scales with $n$). If $n_v$ did not scale with $n$, i.e., if there were an absolute upper bound to the number of ions that can be bound to a vesicle, then for growing $n$ all vesicles would be filled with ions after shorter and shorter time, simply because there more and more unbound ions. Thus, the definition of $w_k$ in \eqref{w_k} as a quantity relative to $n$ is crucial to getting a reasonable hybrid model with good approximation properties. It is very important to note that this scaling assumption does \emph{not} contradict the findings in the biological literature where it is often assumed that vesicles bind maximally 5 calcium ions, with an estimate of $n= 10^2$ ions per vesicle in the spatial domain of interest. There is no contradiction since the limit $n\to\infty$ is a mathematical abstraction used to define a meaningful mean-field limit and \emph{not} biological reality, and since our scaling assumption can be calibrated to agree with the numbers mentioned in the biological literature by setting $a=0.05$ for $n= 10^2$ ions.
\end{rem}

\subsection{Full hybrid model}\label{sec:hybrid}
In the previous section we derived the partial mean-field model in the simplified setting of one vesicle with a fixed position. More generally, from the particle-based dynamics in \sectionname~\ref{sec:particlebased}, one can derive the following partial mean-field model  
\begin{subequations}
\begin{align}
 \dot{c}(x; t) &= \frac{(\sigma^X)^2}{2} \Delta c(x; t) + \sum\limits_{k=1}^m \Bigl(- \ind_{\B(Y_k(t))}(x) \, r^+(w_k(t)) \, c(x; t) + r^-(w_k(t)) \, \mu_k(x) \, a \, w_k(t) \Bigr),\label{Q1} \\
 \dot{w}_k(t) &= r^+(w_k(t)) \, \frac1a \int_{\B(Y_k(t))} c(x; t) \, \di x - r^-(w_k(t)) \, w_k(t) ,\label{Q2}\\
 \di Y_k(t) &= -\biggl(\nabla V(Y_k(t)) + \sum_{\ell\neq k} \nabla U\bigl(Y_k(t) - Y_\ell(t)\bigr) \biggr) \, dt  + \sigma^Y \di B_k^Y\!(t)\label{Q3}
\end{align}
\end{subequations}
for $k\in\{1,\dots,m\}$, where $\mu_k$ defines the distribution of the ion's position after unbinding from vesicle $k$. The model is composed of a PDE~\eqref{Q1} for the concentration of unbound ions~$c$, a collection of ODEs~\eqref{Q2} for the occupancies~$w_k$, and a collection of stochastic differential equations (SDEs)~\eqref{Q3} for the positions of the vesicles~$Y_k$. 

The boundary conditions corresponding to the particle-based dynamics are given by a Neumann no-flux condition $n \cdot \nabla c = 0$ on the domain boundary $\partial\St$ and reflection from  $\partial\St$ for the vesicle positions $Y_k$. It follows, that at all times the distribution of ions (bound or unbound) is conserved:
\begin{equation}\label{conservation}
    \int_\St c(x;t) dx + \sum_{k=1}^m a\,w_k(t)= 1 \quad \forall t\geq 0,
\end{equation}
given that $ \int_\St c(x;0) dx + \sum_{k=1}^m a\,w_k(0)= 1$.

\section{Numerical experiments}\label{sec:NumExp}

For the subsequently discussed numerical experiments we 
employed an Euler-Maruyama discretization of the SDEs~\eqref{position_vesicle} and \eqref{position_ion} to simulate the particle-based dynamics. 
The solution of the PDE~\eqref{Q1} for the hybrid model was approximated by means of a linear-implicit discretization in time and a finite element method in space. For the corresponding ODE~\eqref{Q2} the implicit Euler method was applied, and the SDE~\eqref{Q3} was discretized in time using again the Euler-Maruyama scheme.
It was checked that decreasing time step and grid size yields identical solutions up to sufficient numerical precision.
 
\subsection{Choice of parameter values}\label{sec:baseparameters}
As a base setting, we consider a bounded region~$\St=[0,1]^2\subset \mathbb{R}^2$, as well as  $n= 10^2$ calcium ions, $m=2$ vesicles and $a=0.05$, thus each vesicle has $n_v = 5$ binding sites. For the rate functions we first neglect cooperativity and assume the form  $r^+(w) = \gamma^+ (1-w)$ and  $r^-(w) = \gamma^-$ with $\gamma^+=4$, $\gamma^-=2$, combined with an interaction radius of $\epsilon=0.2$. Later, in \sectionname~\ref{sec:comparisonratefunctions}, we will also consider other rate functions based on cooperativity.

The vesicles move towards the lower domain boundary due to the potential field $V(y)=0.25y^{(2)}$ for $y=(y^{(1)},y^{(2)})\in \mathbb{R}^2$ and are affected by short-range repulsion from other vesicles by the potential $U(y_k-y_\ell) = 0.05\,\exp(-5\norm{y_k-y_\ell})$ for $y_k,y_l\in \mathbb{R}^2$. In all experiments, we assume a noise intensity of $\sigma^X=0.25$ for ions and of  $\sigma^Y=0$ for vesicles. While we want the vesicle dynamics to be stochastic in general, we set $\sigma^Y=0$ in order to make the hybrid model deterministic (solely) to simplify the analysis below significantly. 

\subsection{Comparison of particle-based and hybrid dynamics}\label{sec:basecomparison}
\begin{figure}[h!]
\centering
\begin{subfigure}[h]{0.3\textwidth}
\includegraphics[height=80mm]{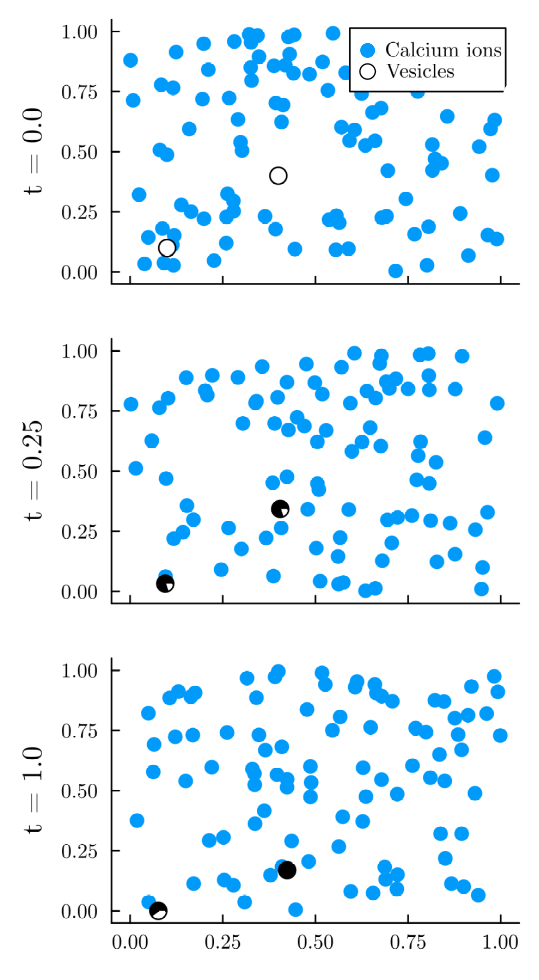}
\caption{\centering Particle-based dynamics \newline (single realization)}
\end{subfigure}
\begin{subfigure}[h]{0.3\textwidth}
\centering
\includegraphics[height=80mm, trim={41 0 0 0}, clip]{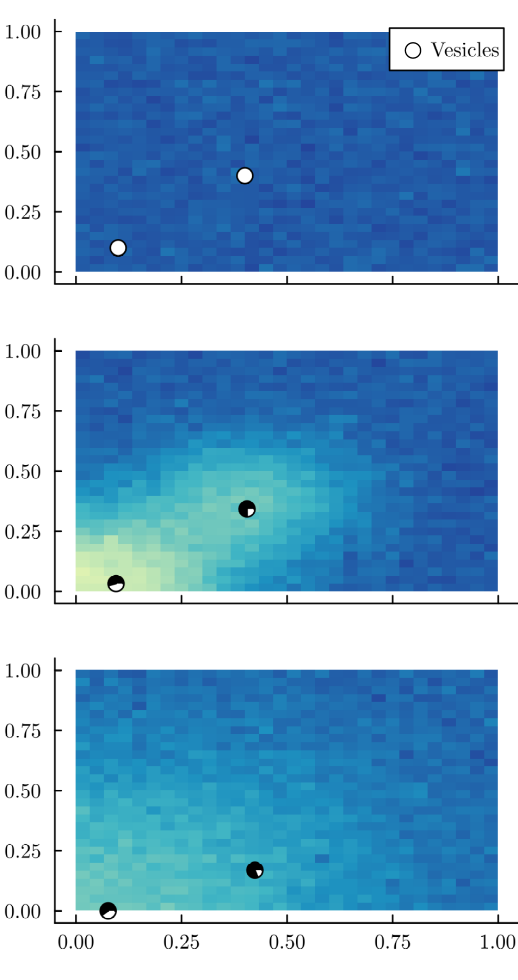}\hspace{15pt}
\caption{\centering Particle-based dynamics \newline (ensemble)}
\end{subfigure}
\begin{subfigure}[h]{0.3\textwidth}
\includegraphics[height=80mm, trim={41 0 0 0}, clip]{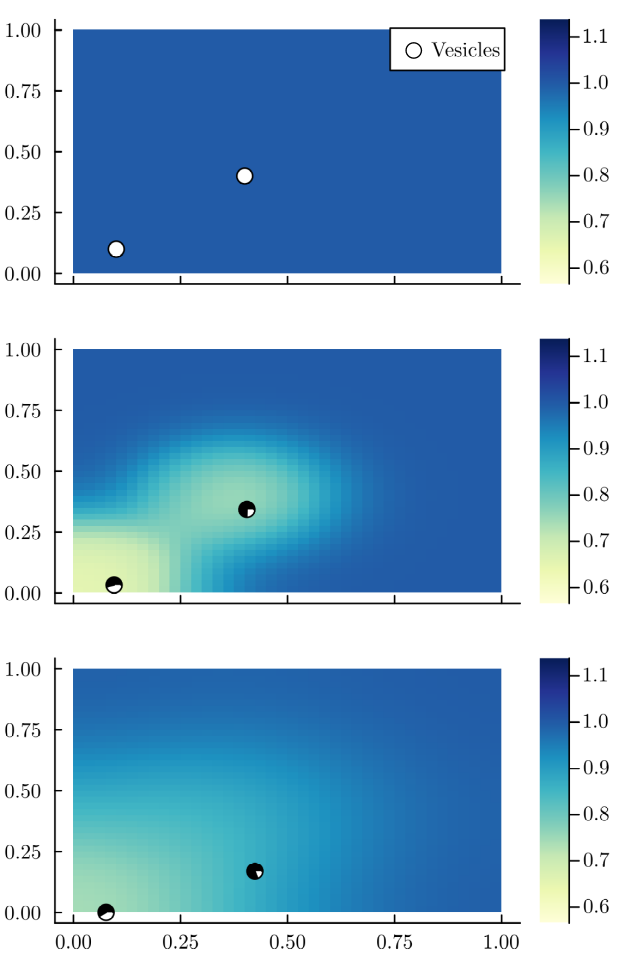}
\caption{Hybrid model  \newline }
\end{subfigure}
\caption{Comparison of the particle-based dynamics and the hybrid model for $m=2$ vesicles and $n= 10^2$ calcium ions, neglecting cooperativity in the rate functions. All other parameter values are specified in \sectionname~\ref{sec:baseparameters}. In panel (a) we depict the positions of all calcium ions in one specific run of the particle-based model, while in panel (b)  the average histogram of calcium ions computed from many runs of the particle-based model is shown with a heatplot. In panel (c) we depict the distribution of calcium ions $c(x;t)$ of the partial mean-field model with a heatplot. The vesicle positions $Y_k$ and occupancies  $w_k$ are visualized with a pie chart. The size of the black pie corresponds to the occupancy of the vesicle. The detailed time evolution of the vesicle occupancies is also given in \figurename~\ref{fig:occupancy}.}
\label{fig:simulations}
\end{figure}

\begin{figure}[h!]
\centering
\begin{subfigure}[b]{0.31\textwidth}
\includegraphics[width=\textwidth]{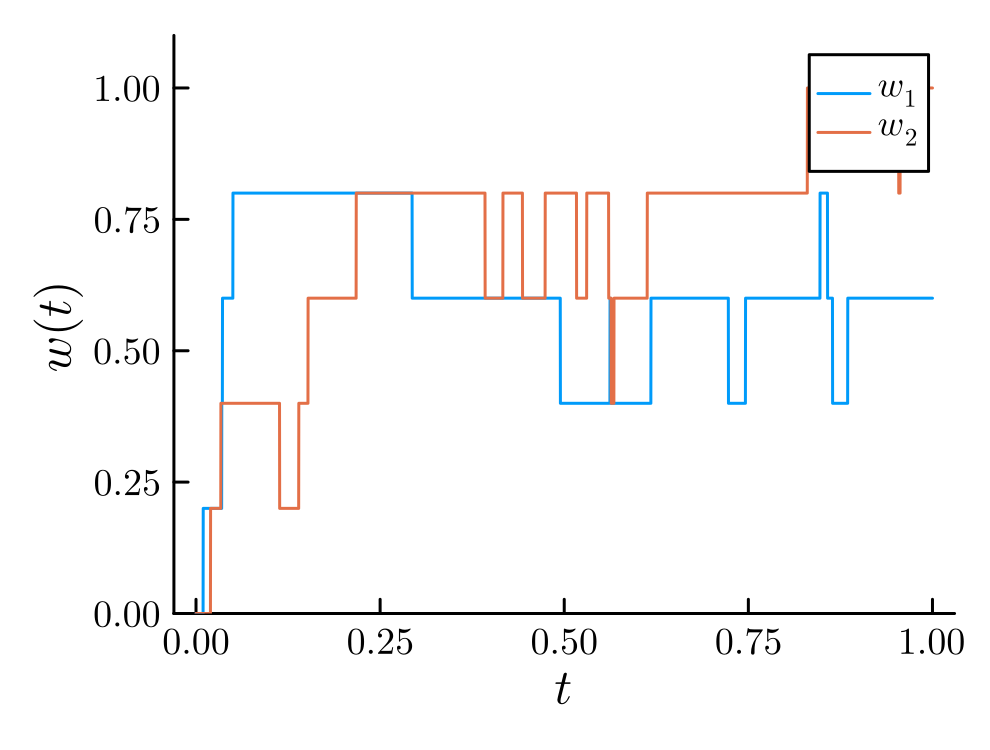}
\caption{Particle-based dynamics \\\centering (single realization)}
\end{subfigure}
\begin{subfigure}[b]{0.29\textwidth}
\includegraphics[width=\textwidth, trim={108 0 0 0},clip]{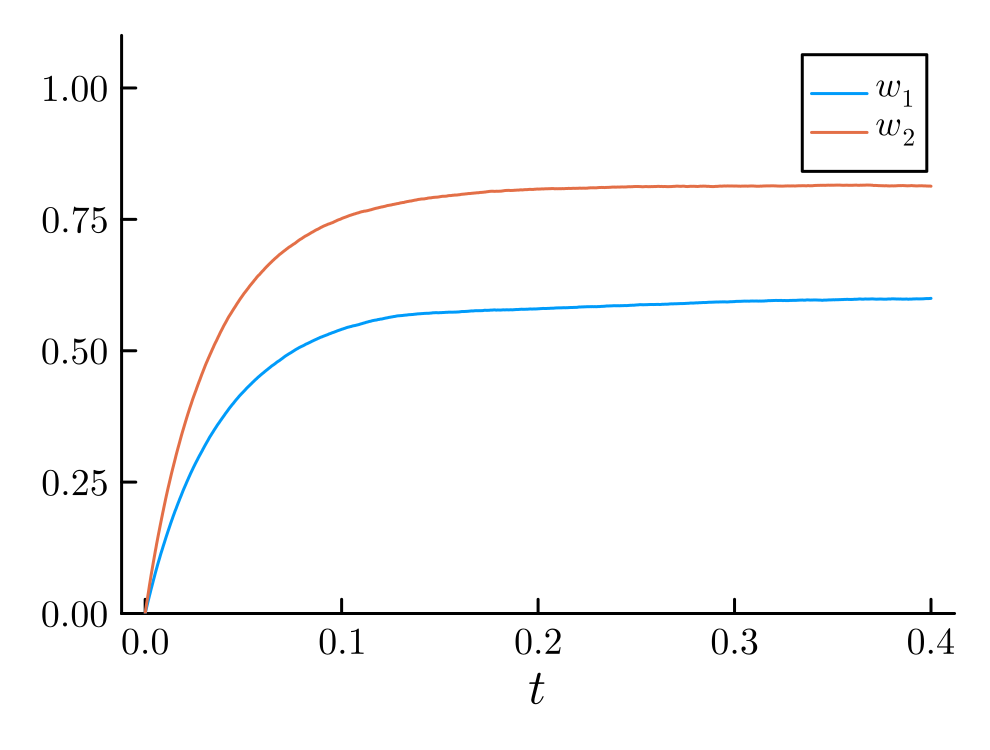}
\caption{\centering Particle-based dynamics \newline (ensemble)}
\end{subfigure}
 \begin{subfigure}[b]{0.29\textwidth}
\includegraphics[width=\textwidth, trim={108 0 0 0},clip]{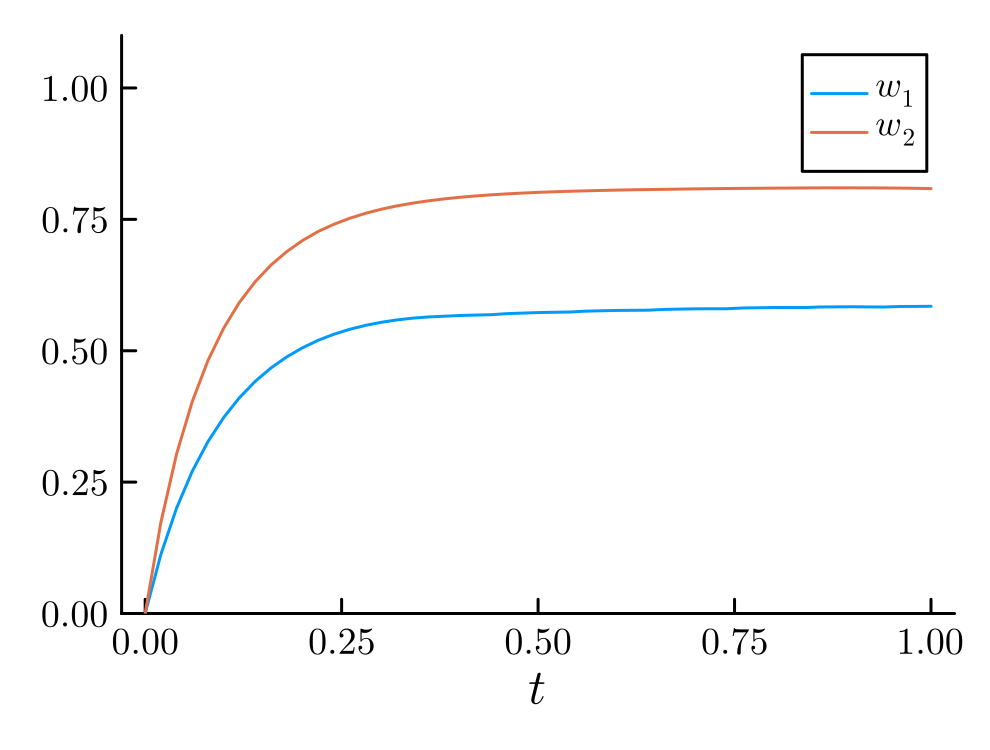}
\caption{\centering Hybrid model \newline}
\end{subfigure}
\caption{Plot of evolution of the occupancy status $w_k(t)$ of the two vesicles $k=1,2$  from \figurename~\ref{fig:simulations} with time $t$. Vesicle $k=1$ corresponds to the vesicle that is initially positioned in the lower left corner of the domain $\St=[0,1]^2$, while $k=2$ corresponds to the vesicle that is initially near the center of $\St$. (a) Plot for a single realization of the particle-based model. (b) Ensemble average over $10^4$ simulations of the particle-based model. (c) Evolution of the continuous occupancy given by the hybrid model.  }
\label{fig:occupancy}
\end{figure}

\figurename{}s~\ref{fig:simulations} and~\ref{fig:occupancy} show the evolution of the spatial distribution of $n= 10^2$ calcium ions and the position and occupancy status for $m=2$ vesicles for both the stochastic particle-based dynamics and the deterministic dynamics given by the hybrid model. In addition, the ensemble average of the particle-based dynamics with respect to $10^4$ simulations is depicted. The parameter values are given in \sectionname~\ref{sec:baseparameters}; the initial positions of the ions were selected randomly from a uniform distribution, and the two vesicles start with occupancy $w_k(t=0)=0$, $k=1,2$. 

For the given choice of parameter values, the hybrid model very well reproduces the average behaviour of the particle-based dynamics. Nonetheless, it is important to note that a single particle-based realization is still highly stochastic and can deviate substantially from the average, which the hybrid model is incapable of capturing. Also note that vesicle $k=1$ has a lower occupancy status than vesicle $k=2$ due to its position. The vesicle's proximity to the domain boundary results in it interacting with fewer calcium ions. Similar parameter values give rise to similarly high approximation quality.

\subsection{Different parameter values and (un-)binding functions}\label{sec:comparisonratefunctions}
In this section we will examine the approximation quality of the average occupancy status of the particle-based dynamics by the dynamics given by the hybrid model when comparing $n= 10^2$ to $n=10^3$. The average quantities for the particle-based dynamics are computed using an ensemble of $5\cdot 10^3$ resp. $5\cdot 10^2$simulations. We will change the form of the rate functions $r^\pm$ to investigate cooperative and non-cooperative behavior as given by the rate functions from \sectionname~\ref{sec:binding} under different values of the rate parameters $\gamma^+$, $\alpha^\pm, \beta$, while fixing $\gamma^-=2$.\footnote{The rate $\gamma^-=2$ may be fixed since by changing $\gamma^+$ we vary the ratio of the binding to unbinding rates.} For simplicity we only consider a single vesicle ($m=1$) and denote $w(t) := w_1(t)$. 

\begin{figure}[h!]
\centering
\includegraphics[width=0.85\textwidth]{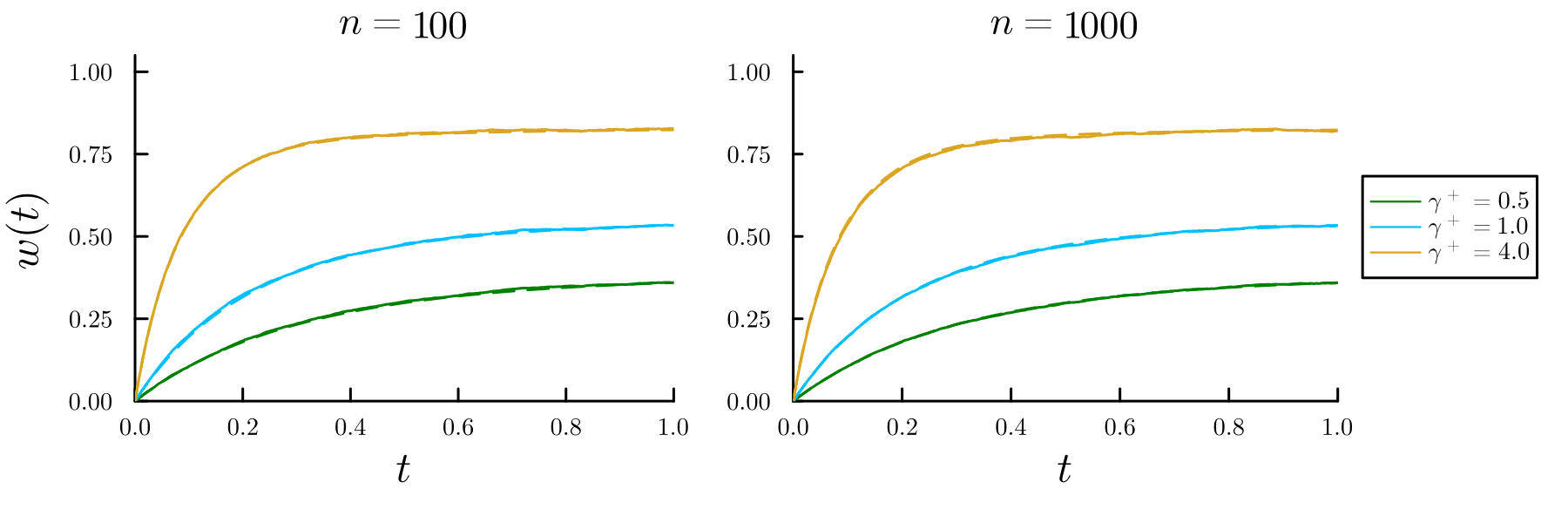}
\caption{Evolution of the occupancy status for the average particle-based dynamics (solid line) and the dynamics given by the hybrid model (dashed line) when binding and unbinding is uncooperative, i.e., $ r^+\?(w) = \gamma^+ \,(1-w) $ and $ r^-\?(w) = \gamma^-$. }
\label{fig:occupancyuncoop}
\end{figure}

\begin{figure}[h!]
\centering
\begin{subfigure}[b]{0.85\textwidth}
\includegraphics[width=\textwidth]{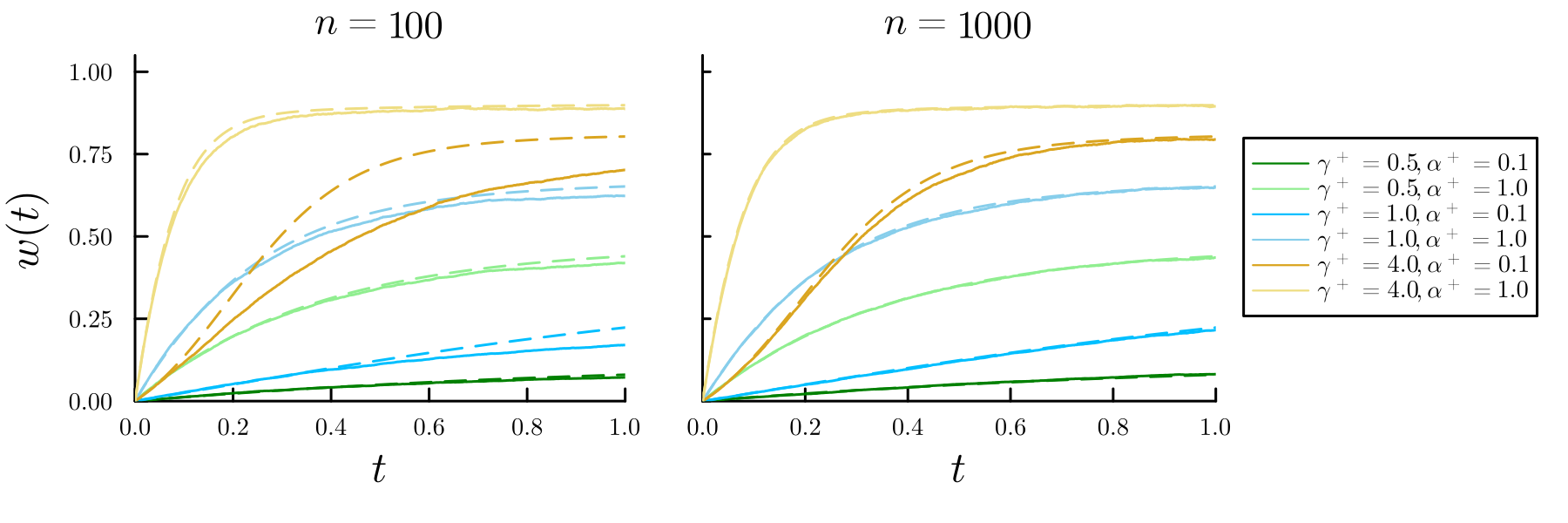}
\caption{Binding is cooperative with $ r^+\?(w) = \gamma^+ \, (w + \alpha^+)(1-w) $, and unbinding is uncooperative  with $ r^-\?(w) = \gamma^-$. }
\end{subfigure}

\begin{subfigure}[b]{0.85\textwidth}
\includegraphics[width=\textwidth]{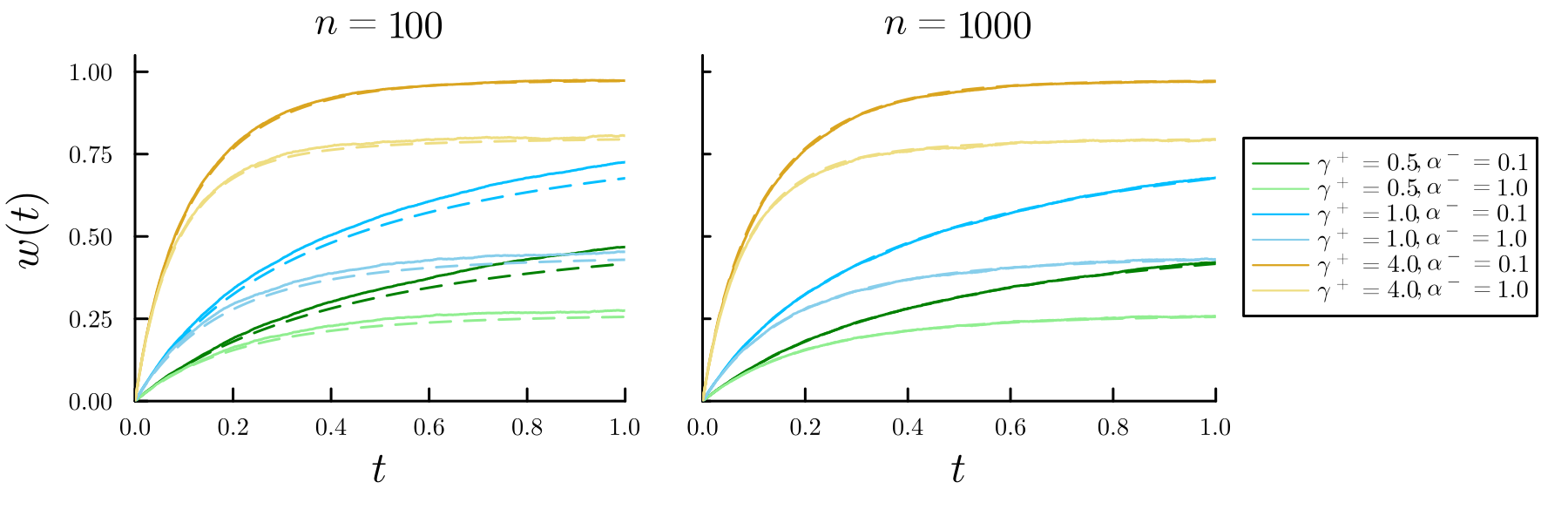}
\caption{Binding is uncooperative with $ r^+\?(w) = \gamma^+ \, (1-w) $, and unbinding is cooperative with $ r^-\?(w) = \gamma^-(1-w+\alpha^-)$. } 
\end{subfigure}

\begin{subfigure}[b]{0.85\textwidth}
\includegraphics[width=\textwidth]{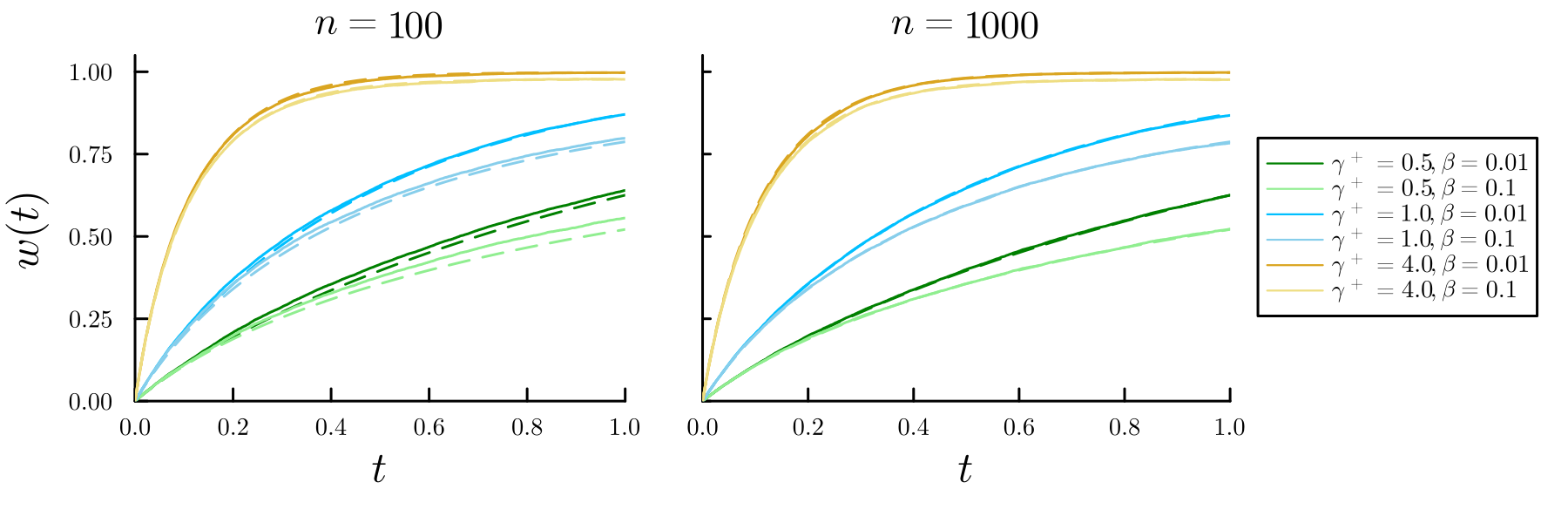}
\caption{Binding is uncooperative with $ r^+\?(w) = \gamma^+ \,(1-w) $, and unbinding is cooperative with $ r^-\?(w) = \gamma^- \beta^{w}$. }\label{fig:coopexp}
\end{subfigure}

\caption{ Evolution of the occupancy status for the average particle-based dynamics (solid line) and the dynamics given by the hybrid model (dashed line) when cooperative binding and/or unbinding is considered.}
\label{fig:occupancycoop}
\end{figure}

When neglecting cooperativity, the approximation quality is already very good for $n= 10^2$ and a wide range of rate parameter values, see \figurename~\ref{fig:occupancyuncoop}. 
In contrast, when binding or unbinding is cooperative, the approximation quality for $n= 10^2$ depends on the specific values of the rate parameters, see \figurename~\ref{fig:occupancycoop}. The approximation is good for some values, while for others, there are discrepancies between the average particle-based dynamics and the hybrid model. However, no clear pattern emerges to explain these discrepancies.  This indicates that, for certain cooperative rate functions and parameter values, a higher number of calcium ions is required for a good approximation between the two models. For $n=10^3$ the approximation quality is high for all tested combinations of rate values. 

\section{Model extensions}\label{sec:Extensions}

The particle-based model introduced above is based on several simplifications and assumptions. First of all, the model is restricted to the pure ion-vesicle binding process. Even regarding this process alone, there are aspects that are not included in the model as it was presented above, e.g., spatial dependence of binding rates and/or noise, effects of charges, buffer proteins, etc. 
Furthermore, the particle-based model ignores many other parts of the neurotransmission process as a whole, like transport through and opening and closing of ion channels, the docking of vesicles to release sites, the recycling of vesicles after release or the neurotransmitter release process itself.

Next, we will shortly outline how the presented particle-based model for the ion-vesicle binding process might be improved. Then, we will show that the model can properly be extended to incorporate ignored parts of the whole neurotransmission process by illustrating how to incorporate
ion transport through an ion channel. 

\subsection{Improved models for the ion-vesicle binding process}

\paragraph{Space-dependent rates and noise.} The particle-based model is based on several specific assumptions about the ion-vesicle binding process. For example, it is assumed above that the binding process happens with equal rate in all of the spatial domain considered. In \cite{kobbersmed2022allosteric}, the authors have postulated that the binding rate is very small away from the membrane, and that additional molecular structures anchored at the membrane may support ion binding. One idea would be to take a non-zero binding rate away from the active zone (smaller than the binding rate at the active zone) and to choose a clearly larger dissociation rate away from the active zone. As soon as such a mathematical model for the spatial dependence of the (un-)binding rate existed, it would be easy to include it into the particle-based and thus also into the partial mean-field model. Furthermore, the diffusion constant might depend on the position of the ions/vesicles. This could be incorporated by making the noise intensity factors in \equationname s~\eqref{position_vesicle} and \eqref{position_ion} position-dependent in the particle-based model. Both improvements would lead to obvious generalization in the partial mean-field model. 

\paragraph{Buffer proteins.} Moreover, in the literature one also finds models that include the reaction of calcium ions with buffer proteins, through which many of the ions in the spatial domain of interest or ions that enter through the calcium channel get bound to buffer proteins and thus only a certain portion of ions eventually reach the vesicles \cite{reva2021first}. Clearly, this could be considered by incorporating an additional species of particles (buffer proteins) with its own diffusive position dynamics and (un-)binding reactions. Consequently, the hybrid model would have to be changed accordingly, e.g., by introduction of an additional PDE for the distribution of buffer proteins. These extensions could be guided by \cite{matveev2022close}, where a deterministic PDE-ODE model of diffusion of calcium ions and reactions with buffer proteins and vesicles is described.

\paragraph{Charges.} Another assumption was to ignore the charge of the ions. While these charges may be screened by different effects within the cellular environment, they should not be ignored completely. Even if we assume that the effect of charge on  binding and unbinding has been considered in the respective rates, there will be an effect on the position dynamics of ions and charge-carrying vesicles.  While the effect of charge on the motion of the vesicles might be modelled by means of additional repulsion or attraction terms in the potential $U$ of \equationname~\eqref{position_vesicle}, the diffusion equation~\eqref{position_ion} would have to be complemented by analogous terms modelling the screened electrostatic repulsion between the ions. A candidate would be
\begin{equation}\label{position_ion_charge}
    dX_i(t) = \delta_{\S_i(t),0} \, \biggl(- \sum_{j \neq i} \nabla \Phi\bigl(X_i(t) - X_j(t)\bigr) \, \di t + \sigma^X \, \di B_i^X\?(t)\biggr), \qquad i \in \{1, \ldots n\},
\end{equation}
where $\Phi$ denotes the screened electrostatic potential of the ions. The introduction of these terms changes the partial mean-field model accordingly, that is, the PDE~\eqref{Q1} for the distribution of unbound ions~$c(x;t)$ gets additional terms and takes the form of a (generalized) Nernst-Planck equation or other electrodiffusion models, cf. \cite{LOPREORE2008,NPN,Sacco2017}: 
\begin{equation}
    \dot{c}(x; t) = \ldots + \nabla \cdot \biggl( c(x; t) \, \nabla \!\! \underbrace{\int_{\St} \Phi(x - x') \, c(x'; t) \, \di x'}_{\text{solution of Poisson's equation}} \biggr) .
\end{equation}
Potentially, one would also have to add analogous terms for the electrostatic interaction between ions and charged vesicles.


\subsection{Adding transport through an ion channel}
The influx of ions through an ion channel can easily by included in both, the particle-based and the partial mean-field model. For the sake of simplicity, we subsequently describe the influx case only. The outflux case can be handled in analogy.

\paragraph{Particle-based model.}
When we want to include an ion channel through which ions can enter the domain, we may model this by assuming that the $n$ ions can not only be (i) unbound and in the domain~$\St$ ($\S_i=0$), (ii) bound and in the domain ($\S_i=k>0$), but also (iii) outside of the domain, denoted by $\S_i=-1$. Then, ions that are outside the domain ($\S_i=-1$) can at a certain rate $k$ (constant rate or time-dependent to ensure a constant inflow number) enter the domain at the channel location~$x_\text{ch}\in\St$, i.e., their position just after entering is given by $X_i=x_\text{ch}\in\St$, or according to a certain distribution centered at $x_\text{ch}$. 
As soon as these ions entered the domain, they are governed by the same rules (diffusion, reactions) as outlined above.

\paragraph{Hybrid model.}
To include an ion channel through which ions can enter the domain, we also model the time-dependent amount of ions outside of the domain, $c_\text{out}(t)$. These ions can enter the domain through the channel at rate $\kappa$, leading to the ODE
\begin{equation}
    \dot{c}_\text{out}(t) = - \kappa \, c_\text{out}(t). 
\end{equation}

In case that the location $x_\text{ch}\in\St$ of the channel lies inside of the domain and not at the boundary, we add the following last term to the PDE~\eqref{Q1}:
\begin{equation}
\dot{c}(x;t) = \frac{\sigma^2}{2} \Delta c(x;t) + \sum_{k=1}^m \left(\dots\right) + \kappa \, c_\text{out}(t) \, p(x), 
\end{equation}
where $p:\St \rightarrow \R^+$ is a non-negative function that integrates to one and determines how the ions enter the domain, e.g. $p(x) = \delta(x-x_\text{ch})$.
Now it holds $\int_\St c(x;t) dx+ \sum_k a\,w_k(t) +c_\text{out}(t)= 1$ for all times $t\geq 0$, in analogy to \eqref{conservation}, assuming that this holds for $t=0$.

Assuming instead that the channel lies on the domain boundary, $x_\text{ch}\in\partial \St$, we replace the Neumann no-flux boundary conditions by $n \cdot \nabla c = \kappa \, c_\text{out} \, p(x)$ on $\partial\St$, where $p:\partial\St\rightarrow \R^+$ is a non-negative function integrating to one along the boundary, e.g., $p(x) = \delta(x-x_\text{ch})$.

\subsection{Further extensions}
Other aspects of the neurotransmission process can be included in similar ways. For example, docking of vesicles to release sites can be integrated by fixing the vesicle position to the membrane/boundary with a certain binding rate upon close contact and starting a new form of position dynamics outside of the boundary for describing neurotransmitter release and diffusion. In the hybrid model, this would lead to an additional PDE for the distribution of neurotransmitter in the spatial domain on the outside of the boundary with addition source terms upon binding of a vesicle to the boundary. 

In conclusion, the particle-based model is flexible enough to allow for incorporation of all aspects of the whole neurotransmission process, as long as good models and parameters (rates, noise intensities, etc.) for the effects to be incorporated become available. The transfer of these additional aspects to the hybrid model then follows the same mathematical recipe as in the derivation above, that is, an (almost) automated derivation process, \emph{except} for the scaling assumptions that have to be made (cf. \remarkname~\ref{rem:occu}).

\section{Conclusion}

This article addresses reaction networks in which spatial and stochastic effects are of crucial importance. For such systems particle-based models allow to describe all microscopic details with high accuracy. However, they suffer from computational inefficiency if particle numbers and density get too large. Alternative models refrain from describing all microscopic details. They reduce the computational effort tremendously by introducing, e.g., a concentration field to represent the particle density, and utilize reaction-diffusion PDEs or similar macroscopic descriptions for the evolution of the concentration field.

The goal of this work is to demonstrate how models on the different resolution levels can be combined into \emph{hybrid models} that seamlessly combine the best of both worlds, describing molecular species with large copy numbers by  macroscopic equations for its concentration field while keeping the stochastic-spatial particle-based resolution level for the low-copy-number species. 

To this end, we introduced a simple particle-based model for the ion-vesicle binding process at the heart of the neurotransmission process. Then, we derived a novel hybrid model and presented numerical experiments that demonstrate that the hybrid model allows for an accurate approximation of the full particle-based model in realistic scenarios. We also discussed how to extend the particle-based model in order to incorporate details and additional aspects of the neurotransmission process presently ignored. It is easy to see how these extensions would results in analogous changes of the hybrid model.

Conclusively, the door is now open to construct hybrid models for other reaction networks with spatial stochastic effects, where one molecular species is only present in low copy numbers in contrast to other high population species. However, the present work also shows that, as usual, the devil is in the details. The form and the approximation properties of the hybrid model crucially depend on the specific scaling properties used. In this work, this is most visible when we revisit the way the upload of ions to one vesicle is modeled: in the particle-based model the ion occupancy of a vesicle is a discrete number; in the hybrid model it becomes a continuous variable that scales with $n$, the number of ions, see \remarkname~\ref{rem:occu}. This kind of scaling assumption will have to be made in every specific case. Further research will have to show which scaling strategies are appropriate for which realistic scenario.

\paragraph{Acknowledgments.}
This work has been partially funded by the Deutsche Forschungsgemeinschaft (DFG) through grant CRC 1114 \emph{Scaling Cascades in Complex Systems} (project no.\ 235221301) and under Germany's Excellence Strategy through grant EXC-2046  \emph{The Berlin Mathematics Research Center} MATH+ (project no.\ 390685689).


\paragraph{Code Availability.} The code is available at \url{github.com/LuzieH/neuro}.
\bibliographystyle{alpha}
\bibliography{ref}

\appendix
\section{Details of the derivation}\label{appendix}
In this appendix, we report the full calculations that bring us from the generator~\eqref{generator} for $(\bX, \bS)$ to the intermediate expression~\eqref{generator_final}. 

We first compute some useful quantities, assuming $\phi,\phi_\ell \in C^{2,0}(\St {\times} \{0, 1\})$ and for $f$ given by~\eqref{observables}:
\begin{subequations}
\begin{align}
    \langle \phi, \rho^n_{\bx, \bs} \rangle ={}& \frac1n \sum\limits_{i=1}^n \phi(x_i, s_i) , \\
    \nabla_{x_i} \langle \phi, \rho^n_{\bx, \bs} \rangle ={}& \frac1n \nabla_1 \phi(x_i, s_i) , \\
    \nabla_{x_i} f(\bx, \bs) ={}& \nabla_{x_i} \psi\bigl(\langle \phi_1, \rho^n_{\bx, \bs} \rangle, ..., \langle \phi_\q, \rho^n_{\bx, \bs} \rangle\bigr) \nonumber \\
    ={}& \frac1n \sum\limits_{\ell=1}^\q \partial_\ell \psi\bigl(\langle \phi_1, \rho^n_{\bx, \bs} \rangle, ..., \langle \phi_\q, \rho^n_{\bx, \bs} \rangle\bigr) \, \nabla_1 \phi_\ell(x_\ell, s_\ell) \label{gradient} , \\
    \Delta_{x_i} f(\bx, \bs) ={}& \Delta_{x_i} \psi\bigl(\langle \phi_1, \rho^n_{\bx, \bs} \rangle, ..., \langle \phi_\q, \rho^n_{\bx, \bs} \rangle\bigr) \nonumber \\
    ={}& \frac1n \sum\limits_{\ell=1}^\q \partial_\ell \psi\bigl(\langle \phi_1, \rho^n_{\bx, \bs} \rangle, ..., \langle \phi_\q, \rho^n_{\bx, \bs} \rangle\bigr) \, \Delta_1 \phi_\ell(x_\ell, s_\ell) \nonumber \\
    &+ \frac{1}{n^2} \sum\limits_{k,\ell=1}^\q  \partial^2_{k \ell} \psi\bigl(\langle \phi_1, \rho^n_{\bx, \bs} \rangle, ..., \langle \phi_\q, \rho^n_{\bx, \bs} \rangle\bigr) \, \nabla_1 \phi_k(x_k, s_k) \cdot \nabla_1 \phi_\ell(x_\ell, s_\ell) \label{Laplacian} , \\
   \MoveEqLeft[8] f\big((x_1, ..., x'_i, ..., x_n), (s_1, ..., 1-s_i, ..., s_n)\big) - f(\bx, \bs) \nonumber \\
   ={}& \psi\biggl( \frac1n \sum\limits_{j=1}^n \phi_1(x_j, s_j) - \frac1n \phi_1(x_i, s_i) + \frac1n \phi_1(x'_i, 1-s_i), ... \biggr) . \label{difference}
\end{align}
\end{subequations}
The identities~\eqref{gradient} and \eqref{Laplacian} follow from the chain rule, and in the final identity~\eqref{difference} we used the trick of removing the contribution of the $i$-th ion with coordinates~$(x_i, s_i)$ from the sum and adding its contribution with the new coordinates~$(x'_i, 1-s_i)$.

Using these formulas, we find \small
\begin{align*}
    \MoveEqLeft[1.82] (L^n f)(\bx, \bs) \\
    ={}& \frac{\sigma^2 }{2n} \sum\limits_{i=1}^n \delta_{s_i, 0} \sum\limits_{\ell=1}^\q \partial_\ell\psi\biggl(\frac1n \sum\limits_{j=1}^n \phi_1(x_j, s_j), ...\biggr) \, \Delta_1\phi_\ell(x_i, s_i) \\
    &+ \frac{\sigma^2 }{2n^2} \sum\limits_{i=1}^n \delta_{s_i, 0} \sum\limits_{k,\ell=1}^\q  \partial^2_{k \ell}\psi\biggl(\frac1n \sum\limits_{j=1}^n \phi_1(x_j, s_j), ...\biggr) \, \nabla_1\phi_k(x_i, s_i) \cdot \nabla_1\phi_\ell(x_i, s_i) \\
    &+ r^+\?(w) \sum\limits_{i=1}^n \ind_{\B(y)}(x_i) \, \delta_{s_i, 0} \biggl[ \psi\Bigl(\frac1n \sum\limits_{j=1}^n \phi_1(x_j, s_j) - \frac1n \phi_1(x_i, s_i) + \frac1n \phi_1(y, 1-s_i), ...\Bigr) - \psi\Bigl(\frac1n \sum\limits_{j=1}^n \phi_1(x_j, s_j), ...\Bigr) \biggr] \\
    &+ r^-\?(w) \sum\limits_{i=1}^n \delta_{s_i, 1} \int_{\mathbb{X}} \mu(\di x') \biggl[ \psi\Bigl(\frac1n \sum\limits_{j=1}^n \phi_1(x_j, s_j) - \frac1n \phi_1(x_i, s_i) + \frac1n \phi_1(x', 1-s_i), ...\Bigr) - \psi\Bigl(\frac1n \sum\limits_{j=1}^n \phi_1(x_j, s_j), ...\Bigr) \biggr] .
\end{align*} \normalsize
To express the generator in terms of the empirical measure, we use the properties
\begin{equation*}
    \sum\limits_{\s' \in \{0, 1\}} \int_{\St} f(x', \s') \, \rho^n_{\bx, \bs}(\di x', \s') = \frac1n \sum\limits_{i=1}^n f(x_i, \s_i) ,
\end{equation*}
which allows us to replace the outermost summations by the corresponding integrals, and
\begin{equation*}
    \qquad w = \frac{1}{\lfloor a \, n \rfloor} \sum\limits_{i=1}^n \delta_{s_i, 1} = \frac{n}{\lfloor a \, n \rfloor} \int_{\St} \rho^n_{\bx, \bs}(\di x', 1) ,
\end{equation*}
by which we shift the meaning of the placeholder~$w$. We then obtain \small
\begin{align*}
    \MoveEqLeft (L^n f)(\bx, \bs) \\
    ={}& \frac{\sigma^2}{2} \sum\limits_{s'\in\{0,1\}} \int_{\mathbb{X}} \rho^n_{\bx, \bs}(\di x', s') \, \delta_{s', 0} \sum\limits_{\ell=1}^\q \partial_\ell\psi\bigl(\langle \phi_1, \rho^n_{\bx, \bs} \rangle, ...\bigr) \, \Delta_1\phi_\ell(x', s') \\
    &+ \frac{\sigma^2}{2n} \sum\limits_{s'\in\{0,1\}} \int_{\mathbb{X}} \rho^n_{\bx, \bs}(\di x', s') \, \delta_{s', 0} \sum\limits_{k=1}^\q \sum\limits_{\ell=1}^\q \partial^2_{k \ell} \psi\bigl(\langle \phi_1, \rho^n_{\bx, \bs} \rangle, ...\bigr) \, \nabla_1 \phi_k(x', s') \cdot \nabla_1 \phi_\ell(x', s') \\
    &+ n \sum\limits_{s'\in\{0,1\}} \int_{\mathbb{X}} \rho^n_{\bx, \bs}(\di x', s') \, \ind_{\B(y)}(x') \, \delta_{s', 0} \, r^+\?(w) \biggl[ \psi\Bigl(\langle \phi_1, \rho^n_{\bx, \bs} \rangle - \frac1n \phi_1(x', s') + \frac1n \phi_1(y, 1-s'), ...\Bigr) - \psi\bigl(\langle \phi_1, \rho^n_{\bx, \bs} \rangle, ...\bigr) \biggr] \\
    &+ n \sum\limits_{s'\in\{0,1\}} \iint_{\mathbb{X}^2} \rho^n_{\bx, \bs}(\di x', s') \, \delta_{s', 1} \, \mu(\di x'') \, r^-\?(w) \biggl[ \psi\Bigl(\langle \phi_1, \rho^n_{\bx, \bs} \rangle - \frac1n \phi_1(x', s') + \frac1n \phi_1(x'', 1-s'), ...\Bigr) - \psi\bigl(\langle \phi_1, \rho^n_{\bx, \bs} \rangle, ...\bigr) \biggr] .
\end{align*} \normalsize
and finally get the more compact expression~\eqref{generator_final} after performing the summations over~$s'$.

\end{document}